\documentstyle[aps,preprint]{revtex}
\newcommand{\de}{\delta}

\newcommand{\gr}{\nabla}

\newcommand{\om}{\omega}

\newcommand{\fr}{\frac}

\newcommand{\Pl}{\partial}

\newcommand{\ts}{\textstyle}

\newcommand{\bee}{\begin{equation}}
\newcommand{\ene}{\end{equation}}
\newcommand{\bem}{\begin{mathletters}}
\newcommand{\enm}{\end{mathletters}}
\newcommand{\bea}{\begin{array}}
\newcommand{\ena}{\end{array}}
\newcommand{\beea}{\begin{eqnarray}}
\newcommand{\enea}{\end{eqnarray}}

\newcommand{\bet}{\begin{tabbing}}
\newcommand{\ent}{\end{tabbing}}
\newcommand{\beb}{\begin{thebibliography}}
\newcommand{\enb}{\end{thebibliography}}

\newcommand{\fder}[2]{\frac{{\ts d \/ #1}}{{\ts d\/ #2}}}
\newcommand{\fpar}[2]{\frac{{\ts \Pl \/ #1}}{{\ts \Pl \/ #2}}}

\begin{document}
\title{Theory of Two Dimensional Mean Field Electron Magnetohydrodynamics }
\author{Amita Das\footnote{Permanent Address:Institute for Plasma Research, 
Gandhinagar India} and P. H. Diamond}
\address{University of California San Diego, La Jolla}
\maketitle 
\begin{abstract}
The theory of mean field electrodynamics for diffusive processes 
in   Electron Magnetohydrodynamic (EMHD) model is presented. 
 In contrast to Magnetohydrodynamics (MHD) the evolution 
of magnetic field here is governed by a nonlinear equation in the magnetic 
field variables.  A detailed description of diffusive 
processes in two dimensions are presented in this paper. In particular, it  
has been  shown  analytically  that the turbulent magnetic 
field diffusivity is suppressed from  naive quasilinear estimates. 
It is shown that for complete  whistlerization of the  spectrum, the turbulent 
diffusivity vanishes.  The question of whistlerization of the turbulent spectrum is 
investigated numerically, and  a reasonable tendency  towards whistlerization 
is observed.  Numerical studies also show the suppression of magnetic field 
diffusivity in accordance with the analytical estimates. \\ 
\vspace{.5in}
 PACS No. 
 \end{abstract}
\section{Introduction}
The transport and amplification properties of a large scale magnetic field 
remains an area of active investigation. This is primarily due to its 
relevance in a variety of physical phenomena. For example,  
 the existence of magnetic field in the universe  is being understood on the  
basis of amplification process by some kind of 
  dynamo mechanism. Another interesting phenomenon is the 
release of high energy bursts in solar flares etc.  It is believed to occur as a 
result of the reconnection of magnetic fields, which can happen  in the presence of 
 finite diffusivity.  However, there is only modest  quantitative understanding 
of these processes.  The amount of 
magnetic energy released by reconnection depends on the value of diffusivity, which  
turns out to be too small to provide an explanation of the vast energy released in 
these bursts. There have been attempts then to understand these phenomenon on the 
basis of turbulent magnetic field diffusivity, which is directly related to the 
question 
of transport of a large scale magnetic field in the presence of turbulence.  
Most theories put forward in these areas are cast within  
the Magnetohydrodynamic system. Lately, however, there has been some work 
which make use of models pertaining to  faster time scales. It is on this regime 
that we are going to focus here. 

 In the present work we address the question of the diffusion  of a  
 long scale magnetic field in the presence of small
scale turbulent magnetic fluctuation ocurring at time scales which are faster 
than the ion response time. For such phenomena the evolution of magnetic field is 
governed by   the electron flow velocity. The ions being stationary, the flow 
velocity of the electrons determines the current and hence is thus directly related 
to the curl of magnetic field. Thus unlike MHD, in this approximation, heretofore 
referred as the Electron Magnetohydrodynamic (EMHD) approximation, the magnetic 
field itself evolves through an explicitly nonlinear equation.  This should be 
contrasted to   the MHD model in which  the nonlinear effects 
creep   indirectly through the lorentz force operating on the plasma flow.       

The paper is organized as follows. In section II we present the salient 
 features of the Electron Magnetohydrodynamics (EMHD) model. In section III we study
 the evolution of mean magnetic field in two dimensions within the framework of 
EMHD description. In two dimensions there is 
no amplification of the large scale field, it can only diffuse. We obtain an expression 
for the effective diffusion coefficient and show that it is suppressed from the naive 
quasilinear estimates. For complete whistlerization, i.e. when the turbulence 
is comprised only of  
 randomly interacting whistler waves  (whistler modes being the normal 
modes of the EMHD model), we show that there is no turbulent contribution to diffusivity. 
This, then raises the pertinent question about  the constituents of the turbulent state 
in this particular model. It becomes important to know whether the 
turbulent   state  comprises entirely of randomly interacting 
whistler waves or  is it merely  a collection of random eddies or is it that a 
 combination of both whistlers and eddies which 
represent the true scenario? We address these question in section IV by numerically 
simulating the decaying turbulence for 
EMHD equations. The initial condition is chosen to be random, i.e. 
no whistlers to begin with. The study of final state reveals evidence of 
 whistlerization.  
In section V we  numerically investigate the problem of diffusion, which shows  
suppression of magnetic field diffusivity, essentially confirming our analytical 
findings   of section III.  Section VI contains the discussion and conclusion.  

\section{the Model}
Electron Magnetohydrodynamics (EMHD) is the theory of the  motion of
magnetized electron fluid in the presence of self consistent and external 
electric and magnetic fields. Such a theory is applicable when the 
time scales of interest are fast (e.g. lying between electron and 
ion gyrofrequencies) so
that ions being massive and unmagnetized play a passive role as a
neutralizing background, and the dominant role in dynamics is played by a 
strongly magnetized electron species. Phenomena having such time scales are often
 encountered in a number of plasma operated devices (e.g. switches, 
focusing devices, fast Z-pinches  etc. \cite{app}). 
Moreover, in the description of collisionless magnetic
reconnection \cite{rec} as well as in certain problems related to ionosphere, 
the EMHD paradigm is invoked frequently. The entire whistler physics is premised on
the EMHD regime of dynamics. 
 
The EMHD model is obtained by using the (i) electron momentum equation 
(ii) the current  expressed in terms of electron velocity  $\vec{J} = 
- n_e e v_e$ as  the ions are stationary at  fast time 
scales depicted  by the model; 
and (iii) the Ampere's law, where displacement current is ignored under the 
assumption ($\om << \om_{pe}^2/\om_{ce}$). The  magnetic field then evolves  
through the following  equation 
\bee
\fpar{}{t}(\gr \times \vec{P}) = \gr \times (\vec{v_e} \times(\gr \times \vec{P})) - 
m_e \nu \gr \times \vec{v_e}
\ene  
Here $m_e$ and $\vec{v_e}$ are the electron mass and the velocity respectively, 
$\vec{P}$ is the canonical momenta defined as  
 $\vec{P} = m_e \vec{v_e} - e \vec{A}/c$ ($\vec{A}$ being the vector potential of 
the magnetic field), $\nu$ represents the electron ion collision frequency. 
Using the current and electron velocity relationship we obtain 
$ \gr \times \vec{P} =e(d_e^2 \gr^2 \vec{B} - \vec{B})/c $; where $d_e = c/\om_{pe}$
is the skin depth. 

It is clear from Eq.1 that the 
$(d_e^2 \gr^2 \vec{B} - \vec{B}) $ is frozen in the electron fluid flow. 
In the limit when the electron inertia can be ignored, it is simply 
the magnetic field which is carried along with the electron fluid. Since  
$v_e \sim - \gr \times \vec{B}$; the evolution equation for magnetic field is 
nonlinear in $\vec{B}$. This can be contrasted with the MHD model 
where the magnetic field 
evolution is governed by an equation which is intrinsically linear in $\vec{B}$. In MHD, 
the nonlinear effects then arise as a result of back reaction on the fluid flow through 
the Lorentz force terms. Basically, in EMHD $\vec{v_e} \sim - \gr \times \vec{B}$, 
and so the  flow 
is directly related to the instantaneous magnetic field; whereas in MHD the evolution of 
flow velocity $\vec{v}$ depends on magnetic field through the Lorentz force 
term and hence $\vec{v}$ has  a memory of the past magnetic field configuration. 
The MHD model is applicable  for scale lengths which are longer than the 
ion skin depth.  EMHD on the other hand depicts phenomenon having scale lengths 
shorter  than the ion skin depth.  
Another  distinction from MHD  arises from the presence of intrinsic scale, viz. 
the electron skin depth $d_e = c/\om_{pe}$ in the EMHD model, which 
separates the two regimes one in which  electron  inertia is important and the other 
where the electron inertia plays no role. The character of the 
EMHD equation changes in these two disparate regimes of scale lengths.       

In two dimensions (i.e. when the variations are confined in $x-y$ plane) 
Eq.1 can be simplified and 
cast in terms of two scalar variables $\psi$ and $b$  which define the 
 total magnetic field by the expression  
$ \vec{B} = {\hat{z}\times{\bf \gr}\psi} + b\hat{z}$. 
 The following coupled set then represents the evolution of these scalar variables  
\bee
\fpar{}{t}(\psi - \gr^2 \psi) + {\hat{z}\times{\bf \gr}b} \cdot \gr (\psi - \gr^2
\psi) = \eta \gr^2  \psi 
\ene  
\bee
\fpar{}{t}(b - \gr^2 b) - {\hat{z}\times{\bf \gr}b} \cdot \gr \gr^2 b  
+ {\hat{z}\times{\bf \gr}\psi} \cdot \gr \gr^2 \psi = \eta \gr^2 b
\ene 
Here we have chosen to normalize length by electron skin depth $d_e = c/\om_{pe}$, 
magnetic field by a typical amplitude $B_0$ and time by the corresponding 
electron gyrofrequency. In the nonresistive limit the above coupled equations support 
the following quadratic invariants 
$$
E = \fr{1}{2}\int[(\gr \psi)^2 + b^2 + (\gr^2 \psi)^2 + (\gr b)^2]dx dy
$$  
which represents the total energy (sum of the magnetic and the kinetic energy),  
$$
H = \int (\psi - \gr ^2 \psi)^2 dx dy
$$
the mean square magnetic potential  and 
$$
K = \int (\psi - \gr ^2 \psi)(b - \gr^2 b) dx dy
$$
the cross helicity. The fields $b$ and $\psi$ are chosen to be uncorrelated 
initially in our numerical simulations. 
On the basis of the existence of these quadratic invariants 
it can be infered that the mean square magnetic potential cascades towards longer 
scale. We will be making use of this later in our derivation for turbulent 
diffusivity. 

Linearizing the evolution equations  in the presence of uniform magnetic field 
$B_0$ pointing in the $y$ direction leads to the following dispersion relation 
$$
\om = \pm \fr{k k_y d_e^2 \om_{ci}}{(1+k^2 d_e^2)}
$$
for whistlers, the normal mode of oscillations in the EMHD regime. It is clear form 
the dispersion relation that the propagation of these waves is preferentially parallel 
to the magnetic field.  Furthermore, the whistler wave excitation leads to the  coupling 
of the form  
$b_k = \pm k \psi_k$ between the two perturbed fields.  This  relation 
between the perturbed fields  then leads to an equipartition 
between the energy associated with  the poloidal and the axial fields. An initial 
unequal distribution of energy in the poloidal and axial fields ultimately has 
a  tendency towards redistribution and achieving equipartition as a 
result of the whistlerization of the spectrum. 
 It is observed  
that time asymptotically the turbulent state in EMHD consists  
 of a gas of whistlers interspersed with a collection of random eddies. 

There has been considerable interest lately to understand features of EMHD turbulence 
both in two and three dimensions in terms of power spectra and the cascade 
properties of the square invariants supported by the model \cite{drbi}. Our attempt here, 
however, is to understand the role of EMHD turbulence in determining the 
diffusion  of long scale magnetic field.

\section{Suppression of turbulent magnetic diffusivity in 2D}

In this section we concentrate on the transport of magnetic field in two dimension. 
In 2D the magnetic field can only diffuse, thus our endeavour here is to estimate 
the effective magnetic diffusivity in the presence of turbulence. 

We will concentrate here on turbulent scale lengths longer than the electron skin depth. 
In this regime of scale lengths i.e. for $kd_e << 1$ the electron inertia effects are 
unimportant and as mentioned in earlier section 
the magnetic field lines are frozen in the electron fluid flow. Thus  turbulence in  
the electron velocity leads to the diffusion of magnetic flux. 
This diffusion of magnetic field lines, arising as a result of turbulence and not due 
to resistivity, is termed as the turbulent diffusivity of the magnetic field. 
The effective 
turbulent diffusivity would thus depend on the electron fluid flow velocity.   
 A naive quasilinear estimate would thus predict that the magnetic field 
diffusivity $\beta \sim \tau v_e^2\sim \tau (\gr b)^2$, where $\tau$ is some averaged 
correlation time for the  electron flow velocity 
$v_e = \hat{z}\times{\bf \gr}b$ in  the $x-y$ plane, and 
$b$ is the $z$ component of the turbulent small scale magnetic field. This suggests 
that the magnetic field diffusion in the $x-y$ plane is solely determined 
by the turbulent properties of the $z$ (i.e. the axial) component of the 
magnetic field. However, this does not represent the complete picture.  
We will now show that the presence 
of small scale turbulence in the poloidal  magnetic field results in the 
suppression of such estimates of diffusivity. This is similar to  
the work carried out by Gruzinov \cite{grpd},  
Cattaneo \cite{cat} and others in the context of 
MHD. In MHD the magnetic field lines are tied to the plasma flow velocity. 
It is observed that the magnetic field diffusivity is suppressed from 
the quasilinear estimates given solely in terms of plasma flow velocity.   
The presence of  small scale turbulence in the magnetic field,  
which opposes the fluid motion through the $\vec{J} \times \vec{B}$ 
backreaction is  found to be responsible for such a suppression.

  We choose to represent the  small scale turbulence in the fields $b$ and 
 $\psi$ as 
$$
b(x,t) = \sum_k b_k(t)exp(i \vec{k} \cdot \vec{r})
$$
$$
\psi(x,t) = \sum_k \psi_k(t)exp(i \vec{k} \cdot \vec{r})
$$
In addition to this we assume the existence of a large scale magnetic field pointing 
along $y$ direction characterized by the magnetic stream function of the following 
form
$$
\psi_0 = \psi_q exp(i q_x x ) + c.c
$$
This  magnetic field has a scale length $q^{-1} >> k^{-1}$ and hence when considering 
averaging over the scale of turbulence this field can be essentially 
treated as a constant in space.  
We are interested in understanding the process of diffusion of this long scale 
field in the presence of small scale turbulence in the variables $b$ and $\psi$, 
i.e. we seek an equation of the kind  
\bee
\fpar{\psi_q}{t} = -\beta q_x^2 \psi_q
\ene
and are interested in determining $\beta$ in terms of the properties of small scale 
turbulence. The $q^{th}$ fourier component of Eq.2 yields 
\bee
(1+q_x^2)\fder{\psi_q}{t} + <{\hat{z}\times{\bf \gr}b} \cdot \gr (\psi - \gr^2
\psi) >_q = -\eta q_x^2 \psi_q 
\ene
The second term in the equation signifies the generation of $q^{th}$ mode as the 
result of nonlinear coupling between the high $k$ turbulent fields. 
The angular brackets indicate the ensemble average. 
The above equation can be rewritten as 
$$
(1+q_x^2)\fder{\psi_q}{t} + i \vec{q} \cdot <{\hat{z}\times{\bf \gr}b} (\psi - \gr^2
\psi) >_q = -\eta q_x^2 \psi_q
$$
We denote $ <{\hat{z}\times{\bf \gr}b} (\psi - \gr^2\psi) >_q$ by $\vec{\Gamma}$ 
representing the nonlinear flux. Since $q_y = 0$, $i \vec{q} \cdot \vec{\Gamma} 
= i q_x \Gamma_x $. The suffix $x$ stands for the $x$ component. Now 
$$
\Gamma_x = <-\fpar{b}{y} (\psi - \gr^2 \psi)>_q = -\sum_k i k_y  (1+k_1^2)
< b_k \psi_{k_1}>
$$ 
where $k_1 = q-k$. 

To estimate the correlation $< b_k \psi_{k_1}>$ we make use of the quasilinear  
approximation where each of these fields gets generated from the other through the 
interaction with the large scale field. 
  Thus we can write
$$ 
<b_k \psi_{k_1}> =  <b_k \de \psi_{k_1} > + <\de b_k \psi_{k_1} >, $$ 
where it is understood that $\de \psi_{k_1}$ is the magnetic perturbation in the plane 
arising as the result of turbulent stretching of the mean magnetic field by the 
electron flow velocity $\hat{z} \times \vec{k} b_k $; and $\de b_k$ is the 
perturbation in the 
elecron flow (viz.$\hat{z} \times \vec{k}\de b_k $) arising from the Lorentz force 
$\hat{z}k_1^2 \psi_{k_1} \times \hat{y} q_x \psi_q$. 
It should be noted 
here that the first term corresponds to that derived from a  kinematic  
treatment, wherein the response 
of magnetic field on flow is not considered. The 
second term  takes account of the   back reaction  of the  magnetic field  
on the  electron velocity. Thus dropping the second term would be  
tantamount to a purely kinematic approximation. 
We will now show that the second term leads to 
a significant suppression of the estimates of diffusivity obtained purely 
from the kinematic treatment. The equations for $\de b_k $ and $\de \psi_{k_1}$ are 
$$
(1+k_1^2)(-i\om_k + \de \om_k) \de \psi_{k_1} = 
- \eta k_1^2 \de \psi_{k_1} - i k_yb_{-k}iq_x(1+q^2)\psi_q
$$
and 
$$
(1+k^2)(-i\om_k + \de \om_k) \de b_{k} = 
- \eta k^2 \de b_{k} - i k_{y1}(k_1^2 - q^2)\psi_{-k_1}i q_x \psi_q
$$
Here $\om$ represents the linear frequency and  $\de \om $ stands for  the eddy 
decorrelation effect arising from the coherent mode coupling. 
Substituting the above expression for $\de b_k$ and 
$\de \psi_{k_1}$ we obtain the following expression for the nonlinear flux
\bee
\Gamma_x = - \sum_k \left( \tau_k ( k_y^2 \mid b_k\mid^2  - 
              k_{1y}^2 k_1^2 \mid \psi_{k_1}\mid^2 )\right)i q_x \psi_q
\ene
where 
$$
\tau_k = \fr{1}{(1+k^2)(-i \om_k + \de \om_k) + \eta k^2}
$$
Here $\tau_k$ represents  the spectral correlation times for the turbulent fields. 
We have  assumed that the turbulent scales are much longer compared to the 
electron skin depth (i.e. $ k << 1 $) in the above derivation. The evolution 
equation for $\psi_q$ under the approximation $q << k << 1$ can then  be written as   
\bee
\fder{\psi_q}{t} = - q_x^2 \left[ \sum_k \tau_k k_y^2
(\mid b_k\mid^2  -  k^2 \mid \psi_{k} \mid^2 )\right] \psi_q - \eta q_x^2 \psi_q 
\ene
The factor inside the square bracket in the right hand side of the above equation 
represents the turbulent contribution to diffusivity. It is made up of two parts. 
The first part, depending on $k_y^2 \mid b_k\mid^2$ represents the kinematic 
contribution and the second part arises as the result of small scale turbulence 
in the poloidal component of magnetic field.  
It is clear that  turbulence in the poloidal component of magnetic 
field contributes towards suppressing the  magnetic field diffusivity. 
It should be noted here that for  complete whistlerization, the spectral components  
of the two fields would be  related  as  $b_k = \pm k \psi_k$, for which the 
turbulent diffusivity 
vanishes exactly.  For this extreme case, diffusion of $\psi_q$ is determined by 
 resistivity alone. It appears then,  that the  understanding of the question of 
whistlerization of the spectrum in the turbulent state is  of paramount 
importance. We will study  this issue in the next section.

We rewrite Eq.7 as  
\begin{eqnarray}
\fder{\psi_q}{t} &=& - q_x^2 \sum_k \tau_k(<v_x^2>_k - k^2<\tilde{B}_x^2>_k) 
\psi_q - \eta q_x^2 \psi_q 
\nonumber \\
             &=& - \fr{q_x^2}{2} \sum_k \tau_k (<v^2>_k - k^2<\tilde{B}^2>_k) \psi_q - 
\eta q_x^2 \psi_q  
\end{eqnarray}
In the above expression  $\tilde{B}_x $ is the $x$ component of the turbulent field.  
In writing the second equality  we have assumed that the turbulence is isotropic. 
Thus we can write 
$$
\beta = \sum_k \fr{\tau_k}{2}(<v^2>_k - k^2<(\gr \psi)^2>_k) + \eta 
$$
The kinematic diffusivity $\beta_0 $ would be just  
$\beta_0 =\sum_k \tau_k v_k^2/2 + \eta $, dependent on the turbulent velocity alone.  
We can then   express $\beta$ in terms of the kinematic 
diffusivity as  $\beta = \beta_0 - \sum_k \tau_k k^2<(\gr \psi)^2>_k/2$. 
Following Gruzinov et al we assume an equivalence of correlation times 
(i.e. assume $\tau_k = \tau$ for each mode ) and 
 write $\beta = \beta_0 -  \tau <k^2> <(\gr \psi)^2>/2$. 
To  estimate $<(\gr \psi)^2>$ we use the  
stationarity  of the mean square magnetic potential. This can be justified on the 
basis of inverse cascade property of the mean square potential. At longer 
scales dissipation due to resistivity is  small and the assumption of stationarity 
of the mean square potential is reasonably good.   
We multiply Eq.2 by $\psi$ and take ensemble 
average. This yields 
$$
<\psi \fder{\psi}{t}> =  \fr{1}{2}<\fder{\psi^2}{t}> = 0
$$ 
$$
<\psi \hat{z} \times \gr b \cdot \gr \psi> = 
\fr{1}{2} \gr \cdot <\hat{z} \times  \gr b \psi^2> = 0 
$$
we thus obtain
$$ 
\eta <(\gr \psi)^2> = B_0 <\psi \fpar{b}{y}> = \beta B_0^2
$$
Substituting for $<(\gr \psi)^2>$ and writing $\tau/2 $ as 
$\beta_0/<v^2> = \beta_0/<(\gr b)^2>$ we obtain 
\bee
\beta = \fr{\beta_0}{1+\fr{<k^2>\beta_0 B_0^2}{\eta <(\gr b)^2>}} = 
\fr{\beta_0}{1+R_m\fr{ <k^2>B_0^2}{<v^2>}}
\ene
Here $R_m$ is the magnetic Reynold's number. It is clear that for $R_m >> 1$ 
the suppression of the magnetic field diffusivity occurs even when the 
turbulent velocity is larger than the effective whistler speed in the presence of  
$B_0$, the magnetic field.
 
\section{Whistlerization}
We have observed in the earlier section that for a turbulent state which 
is  a collection of whistlers alone, the effective turbulent diffusivity 
goes to zero. Thus it is of significance to understand the whistlerization 
of turbulent spectra.   This is identical to studying the question of 
Alfvenization in the context of MHD model. It is interesting to note,  however, 
 that in the MHD model Alfvenization leads to an   equipartition between the  
magnetic and the fluid energies.  However, there can be no 
equipartition between the magnetic and kinetic energies as a result of 
the whistlerization of the spectrum.  Thus, the dominance of 
magnetic or kinetic energies 
 is dependent on  whether  the typical scale of turbulence are longer or shorter  
that the electron skin depth respectively. 
In this paper we have  concentrated on the case where the turbulent scales are 
much longer compared to the electron skin depth. Thus the total energy is 
predominantly magnetic. Whistlerization of the spectrum then leads to  
  an equipartition between the poloidal and the axial field energies.

We seek to understand the question of whistlerization  by carrying out numerical 
simulation. We evolve the two field  $\psi$  and  $b$ by  
  Eq.2 and Eq.3 respectively, using a 
  fully de-aliased pseudospectral scheme. In this scheme the fields $b$ and $\psi$ 
are fourier decomposed. Each of the fourier modes are then evolved, linear part 
exactly, whereas   the nonlinear terms are  calculated in real space and then 
fourier transformed in $k$ space. This requires going back and forth in real and 
$k$ space at each time step. The Fast Fourier Transform (FFT) routines were used to 
go back and forth in the  real and $k$ space at each time integration. 
The time stepping is done using  predictor corrector with the mid 
point leap frog scheme. The simulation was 
carried out  with a resolution of $128X128$ modes as well as  at a 
higher resolution of $256X256$ modes.  
The initial spectrum of the two fields $b$ and $\psi$ was chosen to be concentrated on 
a band of  scales and their phases were taken to be  random. The two fields 
were chosen to be entirely uncorrelated to begin with.

In  Fig.1  we   show a plot  
$\mid b_k \mid $ vs. $ \mid k \psi_k \mid $ for the initial  spectrum. It is clear 
from the figure that the  initial spectrum is totally different from a 
spectrum  whistler waves, which in turn would have shown up in the figure 
as a straight line passing through the origin with unit slope  
basically  depicting the relationship 
$\mid b_k \mid = \mid k \psi_k \mid $ for whistlers.      
In Fig.2 and Fig.3 we  plot for the evolved spectrum 
$\mid b_k \mid $ vs. $ \mid k \psi_k \mid $ for $B_0 = 0 $ and $0.5 $ respectively. 
It is clear that most of the points now cluster close to the origin. It is 
suggestive, when contrasted with the initial condition of Fig.1 that the 
modes are trying to acquire whistler wave relationship. The  scatter in the plot 
indicates  that both eddies and whistlers constitute the final state. 
Thus a   quantitative assessment of the turbulent state as regards  whistlerization 
of the spectra is required. For this purpose we introduce a variable 
\bee
w_k = \fr{abs(\mid b_k \mid^2  - \mid \psi_k \mid^2)}
{(\mid b_k \mid^2  + \mid \psi_k \mid^2)}
\ene
which essentially indicates the fractional deviation of the $k_{th}$ mode from 
being whistlerized. In Table I we list the fraction of modes in the spectrum 
for which $w_k$ is within certain percentage. 
\begin{center}
\bf{TABLE - I}

\vspace{0.5cm}
\begin{tabular}{|c|c|c|c|} \hline \hline
  & \multicolumn{3}{c|}{Fraction of modes Whistlerized} \\ \hline
Permissible  & Initial condition     & Evolved state               & Evolved state\\
\% deviation &                       & $B_0 = 0 $    & $B_0 = 0.5 $ \\ 
\hline  
  2.5        &    0              &   0.028       &   0.031           \\
  5          &    0              &   0.053       &   0.054         \\
  7.5        &    0              &   0.077       &   0.080         \\
  10         &    0              &   0.101       &   0.102         \\ \hline \hline 
\end{tabular}

\end{center}

It is clear from Table I that 
the initial state had zero fraction of modes having deviations, $w_k$ even upto $10 \%$, 
in the final state a reasonable fraction of modes acquire whistlerization within 
a certain percentage of deviation as measured by the parameter $w_k$.  
We also introduce an integral quantity signifying overall whislerization as
 $w = \int w_k dk / \int dk$. For a completely whistlerized spectrum the 
variable $w$ would take a value of  $0$, and  the maximum value that $w$ can have  is 
unity. For our initial spectrum $w = 0.9957$, after evolution (i)for  
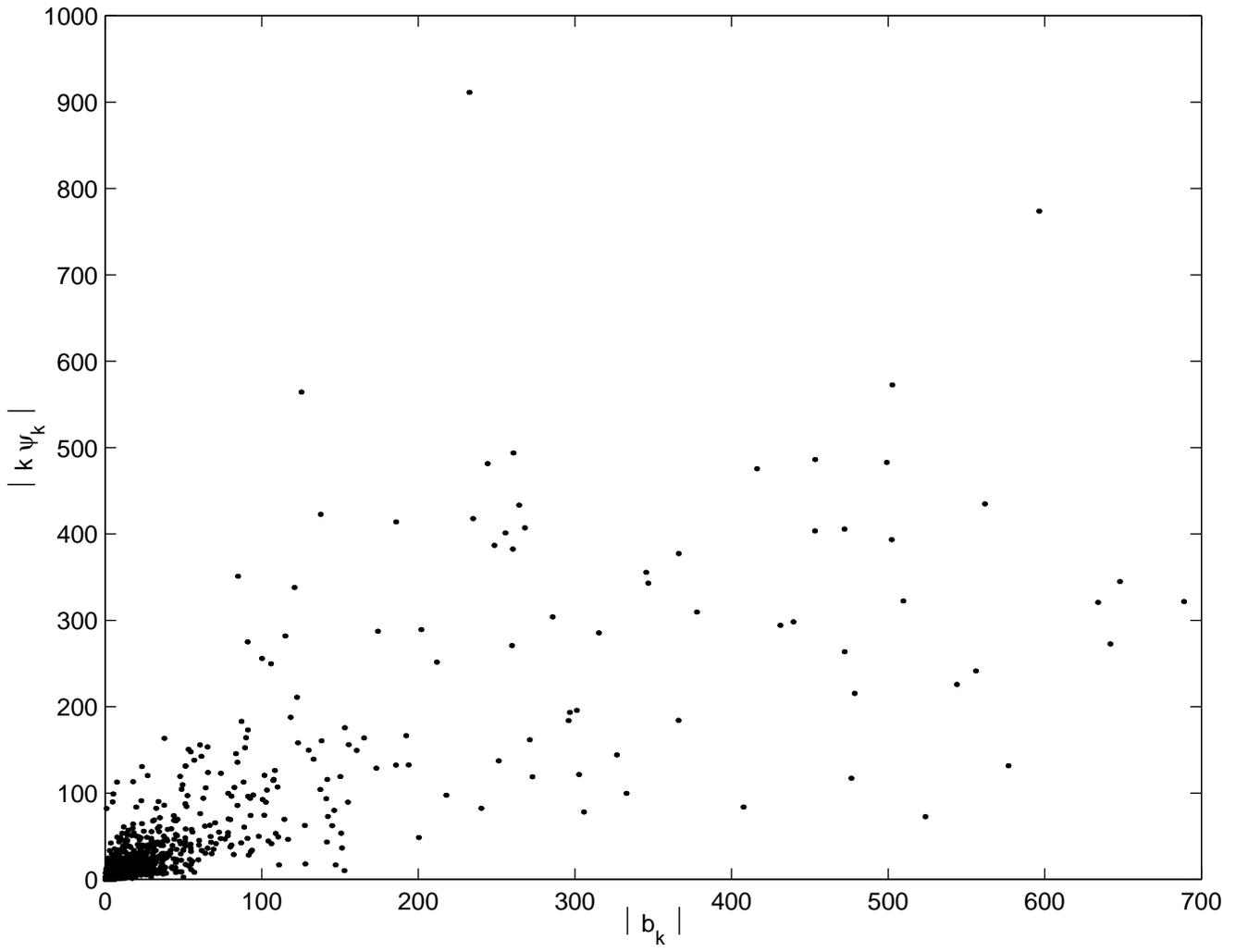
$B_0 = 0$ (corresponding to Fig.1), $w=0.5020$, and (ii) for $B_0 = 0.5$ (Fig.2) 
$w=0.4912$. More  detailed studies of this kind, addressing the 
evolution of whislerization with time (e.g. by studing how $w$ evolves with time), 
its dependence on external magnetic field, etc. are being carried out presently 
and will be presented in a subsequent publication. The question of Alfvenization 
of the spectrum in the context of MHD is also being pursued along similar lines 
and will be presented elsewhere. 

  It is clear from our studies that the whistlerization of the spectrum is not complete. 
Random eddies are also present in the evolved spectrum. This deviation from the 
whistler wave relationship contributes towards the residual effective 
turbulent diffusivity of the magnetic field lines. In the next section we will carry 
out a numerical study to determine the diffusivity of magnetic field 
in the presence of turbulence. 

\section{Numerical results on diffusion }
We saw in  section III that the final expression of the effective diffusivity 
that we obtained was based  on the fact that the effective correlation times 
of the interacting modes were ultimately the same for each of  them. Whether this  
 this really happens can only be verified  by a fully nonlinear numerical 
simulation. We have carried out a set of numerical studies to investigate the 
question of magnetic diffusivity. We observe that the results of our investigation 
agrees with the expression that we have obtained earlier, thereby 
suggesting that the ansatz of local equivalence of correlation time is indeed correct. 
 
The numerical scheme is the same as outlined in the last section. However, in 
addition to evolving the two fields $b$ and $\psi$ 
a number of tracer particles ($N = 1600$) were  placed in the 
two dimensional spatial $x - y$ region of integration. 
The particles were initially placed  uniformly in the $x-y$ plane, 
 and  were then evolved using the Lagrangian electron velocity 
at their location (viz. $\hat{z} \times \gr b$). Since the magnetic field lines are  
tied to the electron flow velocity,  the behaviour of magnetic field diffusivity can be 
discerned  from the diffusion of these particles. Thus the averaged 
mean square displacement 
of these particles is  used as a measure of magnetic diffusivity (e.g. 
$\beta = d<(\delta x)^2>/dt$). This method of evaluating the tracer 
particle diffusivity to study the diffusion of magnetic fields in two dimension  
has been adopted by Cattaneo in the context of the MHD model \cite{cat}.    

It is clear that for  $\eta \ne  0$   and  an initial distribution of 
power with random phases in the various modes for the two fields $b$ and 
$\psi$, Eq.2 and Eq.3 represent the case of 'decaying' EMHD turbulence. 
We refrain from using a 
random stirring force to achieve stationary state as this might lead to  the 
particle displacement being dependent on the characteristics of the random stirrer. 
We will here investigate the case of decaying turbulence and we will present 
results  in the regime 
where the variations can be considered as slow, i.e. we  treat the problem 
in the quasistatic limit.  

   The derivation of our main Eq.9 for the suppression of magnetic field diffusivity 
was premised on the notion of stationarity of the mean square magnetic potential.  
As discussed earlier the cascade of the mean square magnetic potential towards 
longer scales ensures  attaining  such a state. 
 This can be clearly seen  in 
   Fig.4 which shows the evolution  of mean square magnetic potential with time. 
It is clear that the percentage  variation in $\int \psi^2 dx dy$ is small after 
$t = 200$. For the purpose of our 
calculations  in  all our  numerical runs we have restricted 
to the region where the percentage variations in $\int \psi^2 dx dy$ is below $3 \%$.  

 In Fig.5 we show the mean square displacement of the tracer particles with 
time. The thick line indicated by the label 'kinematic'  essentially corresponds to the 
displacement when the uniform magnetic field in the $y$ direction  $B_0 $ is 
chosen to be zero. We 
will designate the slope of this curve as $\beta_{kin}$, the kinematic diffusivity. 
The other two lines essentially correspond to  the longitudinal and the transverse 
displacement in the presence of a uniform magnetic field $B_0 = 1$ 
along the $y$ diection. 
It is clear from the figure that the slope of the kinematic curve is larger than the 
other two curves which correspond to the displacement for finite  $B_0$. 
This clearly indicates  that the presence of $B_0$ suppresses the diffusivity;  
the conclusion we arrived at earlier in the last section. However, 
 longitudinal displacements of the tracer particles are larger compared to their 
transverse displacement, suggesting that the assumption of isotropic turbulence 
in not valid in the presence of uniform magnetic field. There has been indications 
in earlier works both in MHD \cite{sheb} as well as in EMHD \cite{icpp} 
that the presence of strong magnetic 
field results in anisotropy of the spectrum. Our results showing distinct values for   
 the longitudinal and the transverse diffusivity is another evidence for anisotropic 
turbulence in the presence of external magnetic field.   

We next investigate the question whether the supression of diffusivity 
 with increasing  magnetic field is indeed given by the kind of expression (Eq.9) that 
we have obtained in the earlier section. For this purpose we carry out several 
numerical runs with varying strength of the   magnetic field. 
The diffusivity $\beta$ for each case is then given by the slope of the displacement 
of the tracer particles. It is clear from Fig.5 that the curve is jagged, essentially 
signifying that  $\beta$, the diffusivity estimated from the slope  of such a curve 
is a  statistical quantity. We take a time average given by 
$$
\beta(t_2-t_1) = \fr{1}{t_2-t_1} \int_{t_1}^{t_2} \beta(t) dt
$$ 
The choice of $t_2-t_1$ is such that the in this duration the turbulence can essentially be treated as quasistationary. The averaging procedure eliminates the 
statistical fluctuation in the estimate of diffusity and it is observed that with 
varying $t_2$ the slope  asymptotes to a constant value for each case. 
      
In  Fig.6  the $y$  axis represents  $\beta_{kin}/\beta$ and along the $x$ axis 
we vary $B_0^2$. It is clear from the plot that the data points nicely 
fit  a straight line, as our  analytical expression predicts. 

\section{Discussion}
There are two important results  of our present work. First, we have been able to 
show that the turbulent EMHD state shows tendencies towards  whistlerization. 
The spectrum is only partially whistlerized, suggesting that  
both eddies and randomly interacting whistlers constitute the turbulent state. 
Secondly, we have carried out studies to understand the diffusion of long scale 
magnetic field in the context of Electron Magnetohydrodynamics. We have shown that 
the effective  diffusivity due to turbulence in the electron flow velocity gets 
suppressed in the presence of small scale turbulence of the magnetic field. For 
complete whistlerization the turbulent diffusivity vanishes. However, since the 
turbulent state is only partially whistlerized the effective diffusivity does not 
vanish it only gets  suppressed  from pure kinematic estimates. We have confirmed 
these results numerically. 

  The problem of diffusion of magnetic field lines is of great interest, as 
it provides a mechanism for the reconnection of magnetic field lines which 
is thought to underlie an understanding of the rapid release of energy 
in several solar and astrophysical contexts. The resistive diffusion turns out 
to be too small to explain the large amount of energy released. This had 
initiated efforts towards understanding the phenomenon of turbulent diffusivity 
of magnetic field lines. Earlier attempts on this were based on the  
Magnetohydrodynamic approximation. However, it was shown theoretically by 
Gruzinov et al \cite{grpd} and    numerically by Cattaneo \cite{cat} that the 
value of turbulent diffusivity gets suppressed in the presence of turbulence 
in small scale magnetic field. Recently, attempts to understand the reconnection 
phenomenon in the context of Electron Magnetohydrodynamics are being made 
\cite{rec}. Our 
work in this context becomes relevant, as we have shown here that the naive 
quasilinear estimates do not provide a complete picture. The effective  diffusivity 
gets suppressed in the presence of turbulence in the magnetic field, with whistlerization 
of the spectrum playing an important role in this regard. 

Other  issue that we would like to point out  in this regard is the role of 
whistlers in  EMHD turbulence. Some recent studies  on EMHD turbulence 
categorically rule out the presence of whistler effect in determining the 
energy tranfer rate on the basis of the numerically observed scaling of the 
power spectrum \cite{drbi}.  We have, on the other hand  shown here that there is a 
 tendency towards whistlerization of the turbulent spectra and that 
directly influences the effective diffusivity of the magnetic field lines. 
Invoking the Prandtl mixing length argument, which relates the transfer rate 
to the effective diffusivity, the question of whistler effect being present or not 
remains debatable. Moreover, we also have evidence of  
  anisotropization of the turbulent spectrum in the 
presence of external magnetic field ( this work will be presented elsewhere)
which further points towards a subtle role of whistlers in governing the EMHD 
turbulence. \\
{\bf Aknowledgement}: We would like to thank the  San Diego 
Supercomputer centre, an NSF funded site of NPACI for providing 
 computing time  on T90 supercomputer for this work.  This 
research was supported by DOE Grant No. DE-FG03-88ER-53275.

\newpage
\begin{center}
FIGURE CAPTION
\end{center}
\vspace{0.2 in}
\begin{itemize}
\item[Figure 1]
Plot of $\mid b_k \mid$ vs. $\mid k \psi_k \mid$ for the initial spectrum. 
\item[Figure 2]
Plot of $\mid b_k \mid$ vs. $\mid k \psi_k \mid$ for the evolved spectrum when the 
external field $B_0 = 0$.
\item[Figure 3]
Plot of $\mid b_k \mid$ vs. $\mid k \psi_k \mid$ for the evolved spectrum when the 
external field $B_0 = 0.5$.
\item[Figure 4]
Evolution of mean square magnetic potential. 
\item[Figure 5]
 Mean square displacement of the tracer particles with time is shown, thick lines 
(kinematic) shows the displacement in the absence of any external field. The 
other two lines indicated by  'longitudinal' and the 'transverse' show the 
 mean square displacement of the tracer particles along and across the external 
magnetic field $B_0 = 1$. 
\item[Figure 6]
A plot of $\beta_{kin}/\beta$ vs. $B_0^2$.   
\end{itemize}   
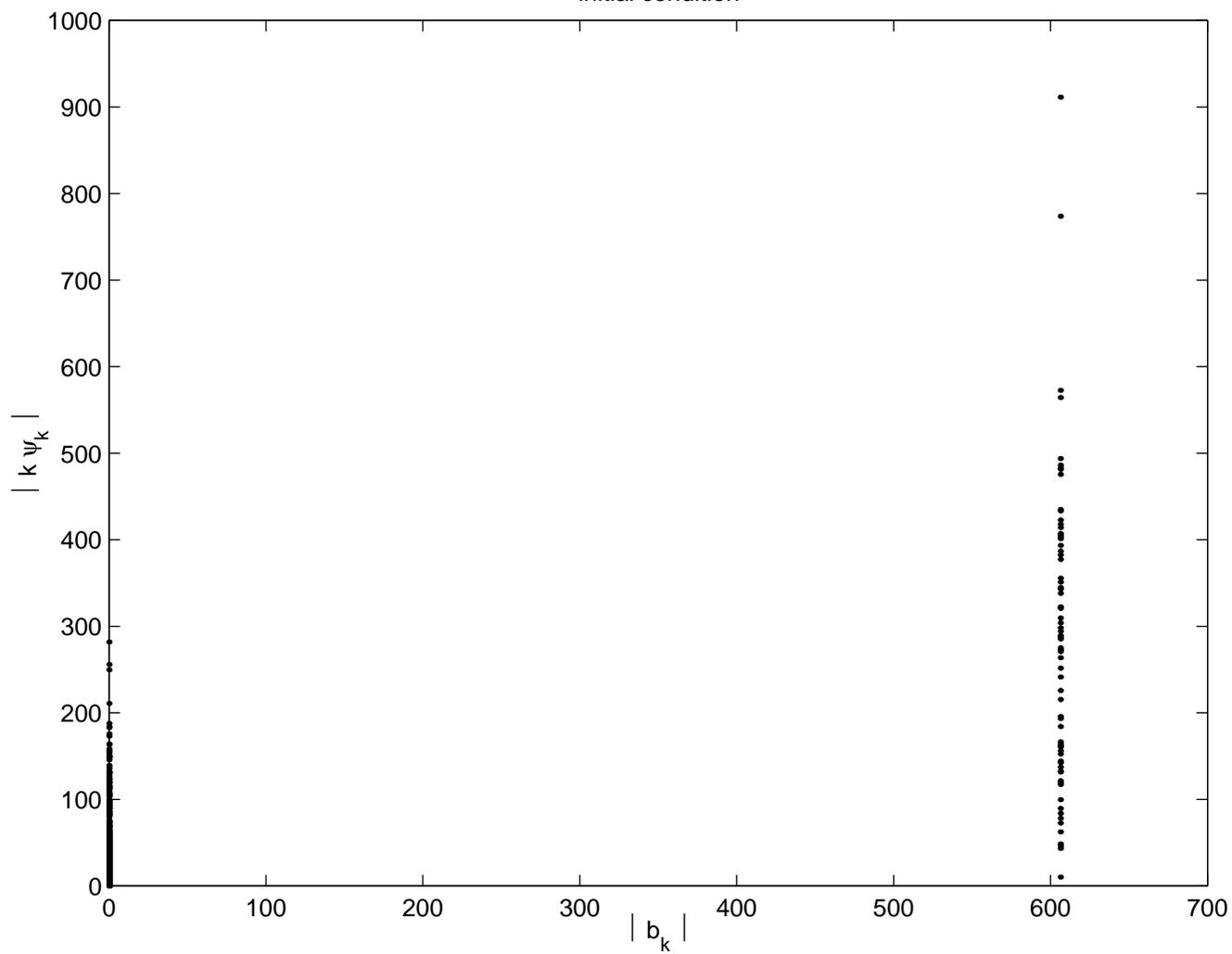
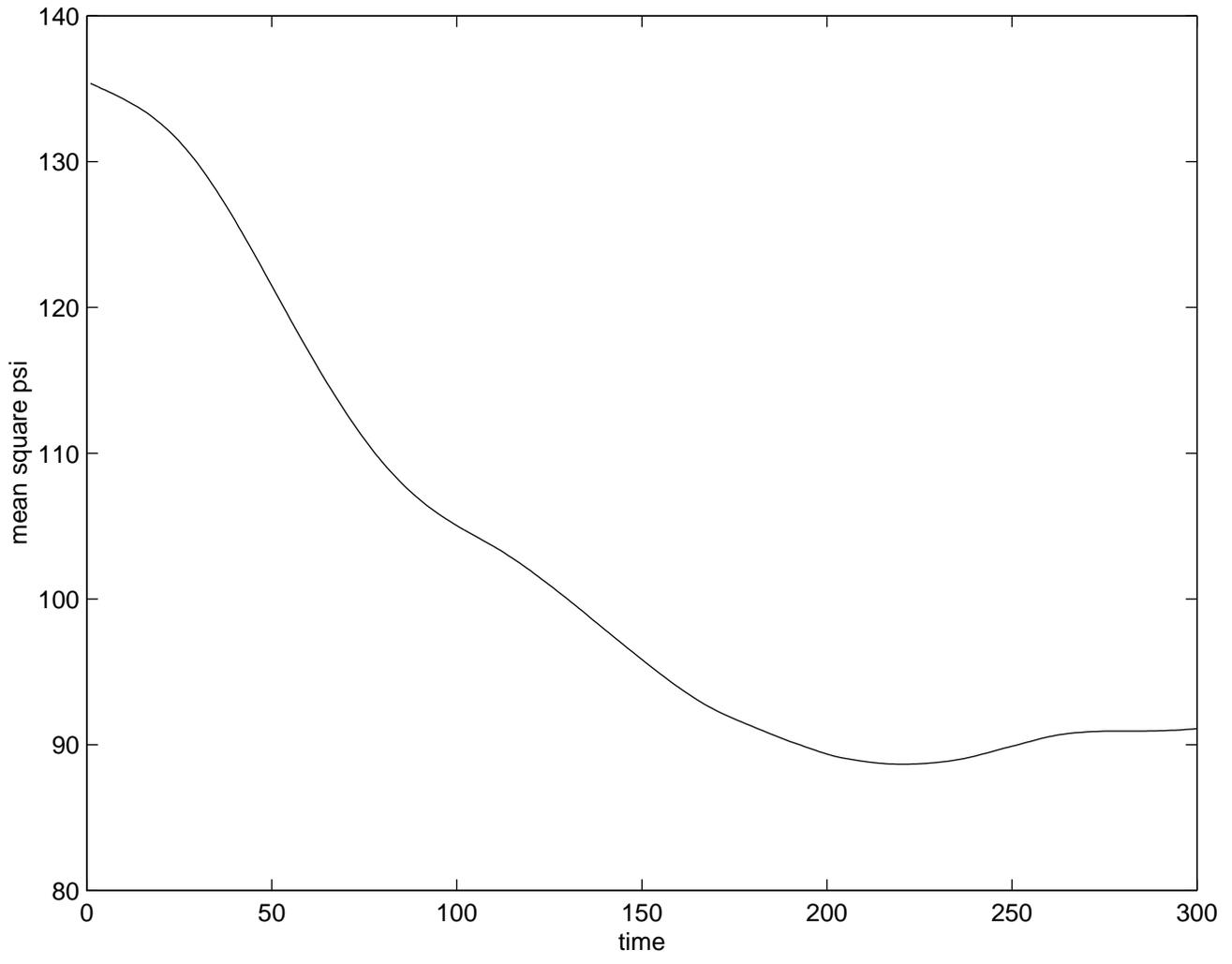
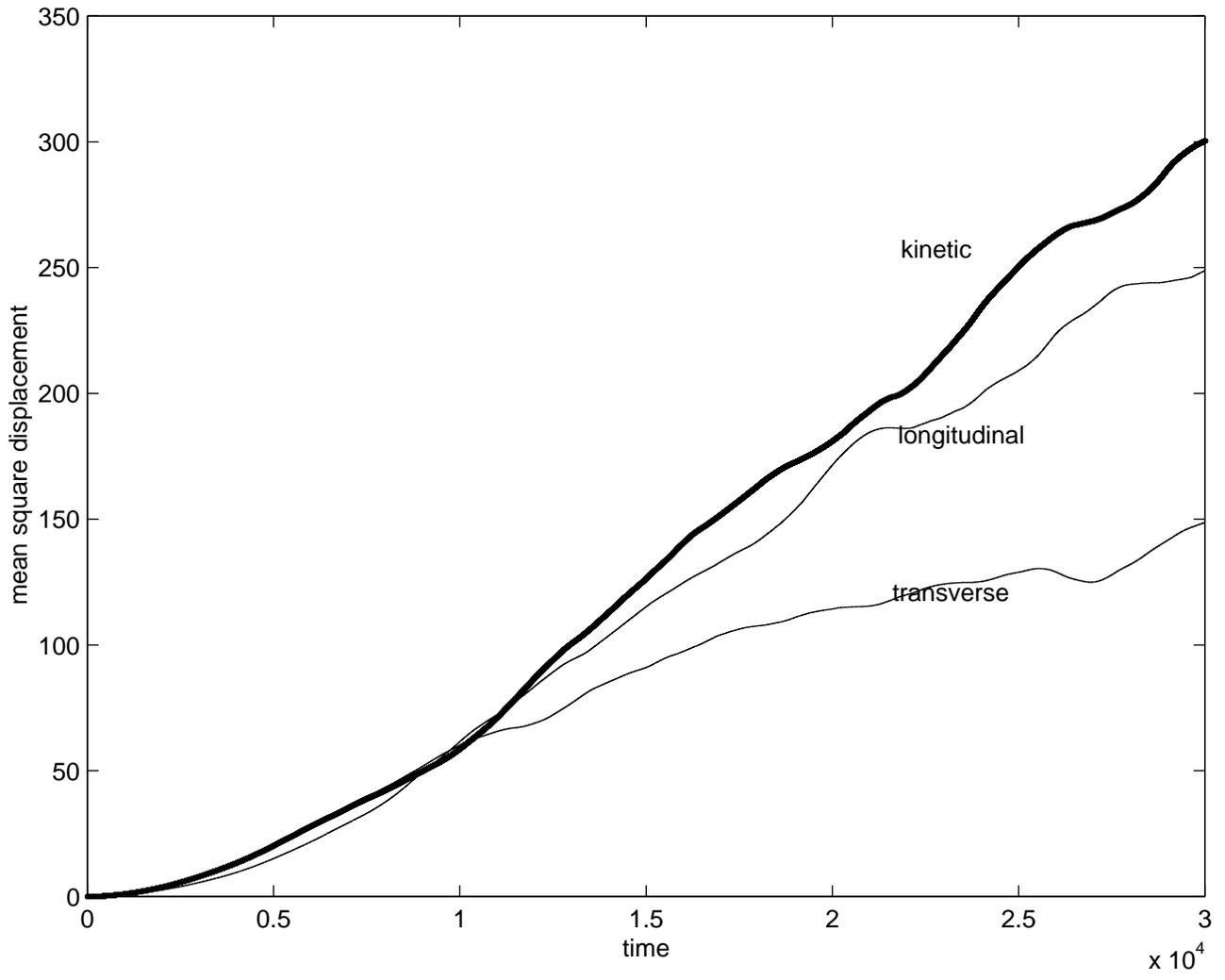
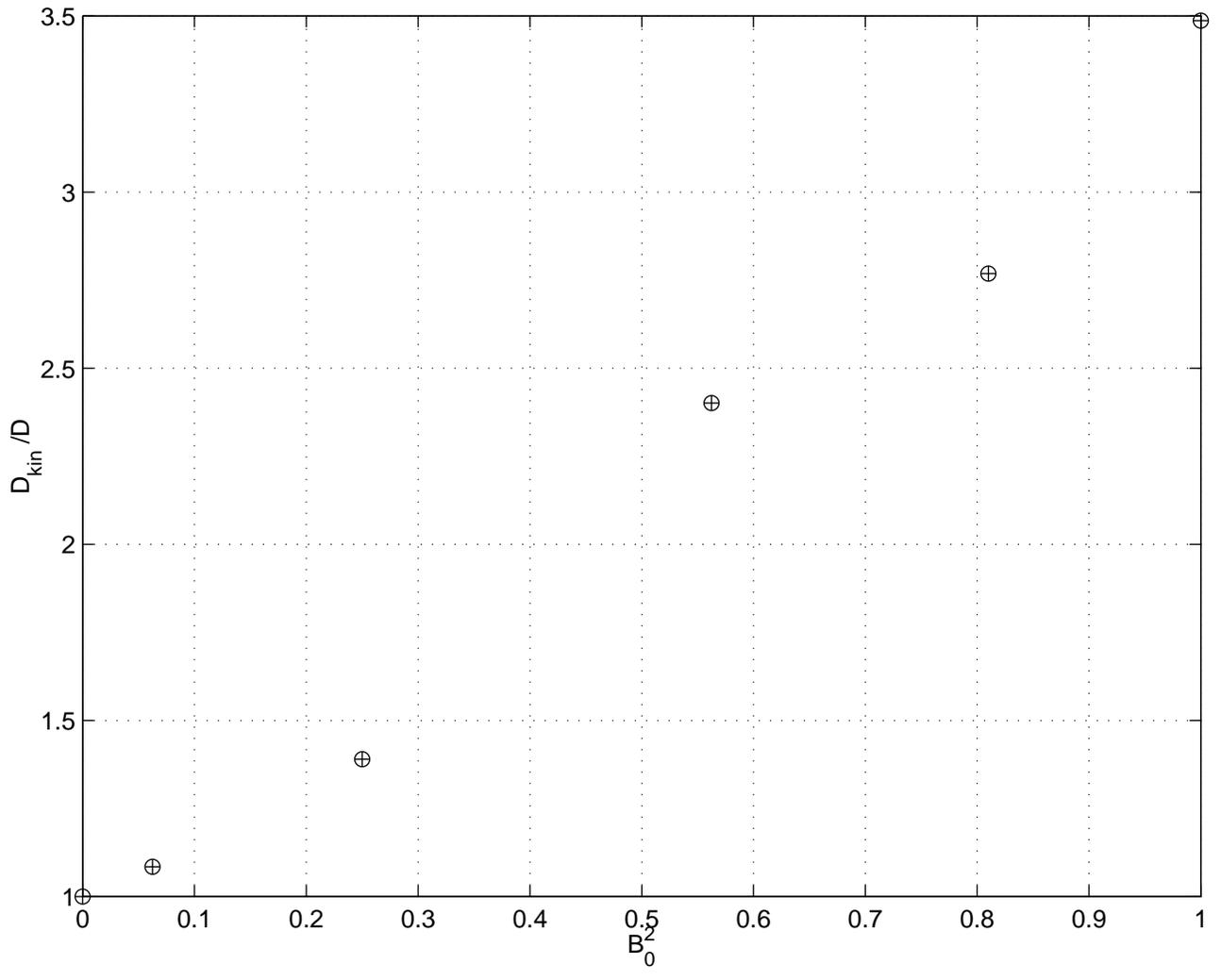
\end{document}